\documentclass[12pt]{article}
\usepackage{epsfig,subfigure,color}
\usepackage{authordate1-4}

\begin{document}
\title{The recurrence time of Dansgaard-Oeschger events and 
limits on the possible periodic component}

\author{Peter D. Ditlevsen, Mikkel S. Kristensen and Katrine K. Andersen\\
The Niels Bohr Institute, Department of Geophysics,\\
University of Copenhagen, Juliane Maries Vej 30,\\
 DK-2100 Copenhagen O, Denmark.\\
}

\date{\today}
\maketitle
{\bf By comparing the high-resolution isotopic records from the GRIP and NGRIP 
icecores, we approximately separate the climate signal from local noise to
obtain an objective criterion for defining Dansgaard-Oeschger events. Our analysis identifies 
several additional short lasting events, increasing the total number of DO events to 27 in the period 12-90 kyr BP. 
The quasi-regular occurrence of the DO events could indicate a stochastic or coherent resonance
mechanism governing their origin. From the distribution of waiting times we obtain a statistical upper 
bound on the strength of a possible periodic forcing. This finding indicates that the climate shifts 
are purely noise driven with no underlying periodicity.   
}

\section*{Introduction}
Abrupt temperature shifts between cold (stadial) states and  
warm (interstadial) states (Dansgaard-Oeschger
events)
are observed in the Greenland icecore isotope 
records \cite{dansgaard:1993,grootes:1993}. 
The transitions into these interstadials were 
abrupt and with temperature changes
estimated from the paleo-record of the order of 10-15 $^oC$ occuring within decades. 
These two distinct quasi-stable climate 
states are most likely linked to different modes of the Atlantic thermohaline circulation (THC)
\cite{broecker:1985,stocker:1992}. The circulation in the warm state was similar to the present 
interglacial state, while in the cold state the sinking took place at lower latitudes and to 
shallower depths \cite{alley:1999,ganopolski:2001}. 
A complete cessation of deep water formation is indicated by some models \cite{schmittner:2002}.

Identification of the mechanism causing these climate shifts 
is of primary importance for understanding the stability and mode
of operation of this component of the climate system. Comparing the high-resolution
isotope records from GRIP and NGRIP \cite{ngrip:2004}, we can separate the climate
signal from the local variability and the glaciological noise in the icecores. We observe that for
timescales longer than approximately 30 years the two icecores are strongly
correlated. Using the 30-year averaged record we thus obtain an objective measure 
to define the DO events. The initiations and terminations
of the DO events are defined from consecutive up-crossings and down-crossings through
two levels in the anomaly record. The definition of the DO events by this procedure is
quite robust with respect to the choice of anomaly levels. Comparing to the original
visual numbering of the DO events \cite{dansgaard:1993} we find several additional isotopic fluctuations 
that qualify as DO events. For example, DO event 2 consists of two
closely spaced DO events, which based on the stratigraphy devised by Walker et al.
\shortcite{walker:1999} and Bj{\"o}rk et al. \shortcite{bjork:1998}, 
becomes Greenland Interstadial 'GI2a' (youngest) and 'GI2c' (oldest)
separated by the stadial state 'GI2b'. Using these conventions the succeeding period is the
Greenland Stadial 'GS2'. For further explanation see the figure caption (figure \ref{newDO}).      

The isotope record
from the GISP2 icecore, dated using stratigraphic methods \cite{meese:1997},  shows a significant peak 
in the
spectrum at 1470 years \cite{yiou:1997}, indicating 
a possible periodic forcing of the climate 
system. 
Alley et al. \shortcite{alley:2001}
proposed this to be a stochastic resonance while Timmermann et al. \shortcite{timmermann:2003}
proposed a coherence 
resonance phenomenon. The climate system itself is 
dominated
by strongly fluctuating, irregular, fast timescale noise, so it is highly 
unlikely that 
a strictly periodic signal would be internally generated. Only simplified 
models with few degrees of freedom exhibit strict cyclic behavior \cite{paillard:1994}.
In a recent study Roe and Steig \shortcite{roe:2004} notes that, 1470 yrs periodicity aside,
the Antarctic Byrd record 
is comparable with a simple autoregressive process and the same is true for the 
Greenland records if an additional simple threshold rule is imposed. 

A periodic component in the signal
would likely be the result of a non-linear response to a weak external 
periodic forcing, though the origin of such a forcing has not yet been identified.       
By observing the waiting times between consecutive DO events it was noted that 
the record could be interpreted as having a preferred waiting time of 1470 years 
and
multiples of this period, corresponding to the system skipping a few transitions 
but still 
switching in phase with the external periodicity \cite{alley:2001,schulz:2002,rahmstorf:2003}. Observing the waiting times 
rather than the
power spectrum is advantageous in a noisy signal where the cyclicity is far from being 
sinusoidal.
In this case a large portion of the power will be in overtones, which possibly 
brings the 
power in the peak below detection into the noise level. This is the case for the 
icecore 
signal where the transitions into the interstadials are rapid and the transitions into
the stadials are more gradual.   

By comparing the icecore record with stochastic resonance models we are able to 
estimate an upper bound on the strength of
an external periodic forcing potential in comparison to the internal barrier between the two climate
states. We find that with maximum likelihood the data are not a result of a process with a periodic component. 
Since there still remains discrepancy between the GRIP and the GISP2 datings we have
performed our analysis on both timescales. 
The GISP2 timescale agrees well with the U/Th dating of the Chinese Hulu Cave stalagmite record \cite{wang:2001}
while the dating of the French Villars Cave stalagmite record \cite{genty:2003} is somewhere in between. 
While we
cannot reject the case of no periodic component we can determine an upper limit 
of a periodic forcing such that a strength above this value can be rejected with 
98\% confidence. This upper limit varies slightly between the GRIP and the GISP2 timescales.

\section{Defining the Dansgaard-Oeschger events}
\begin{figure}[!H]
\begin{center}
\epsfxsize=85mm 
\epsffile{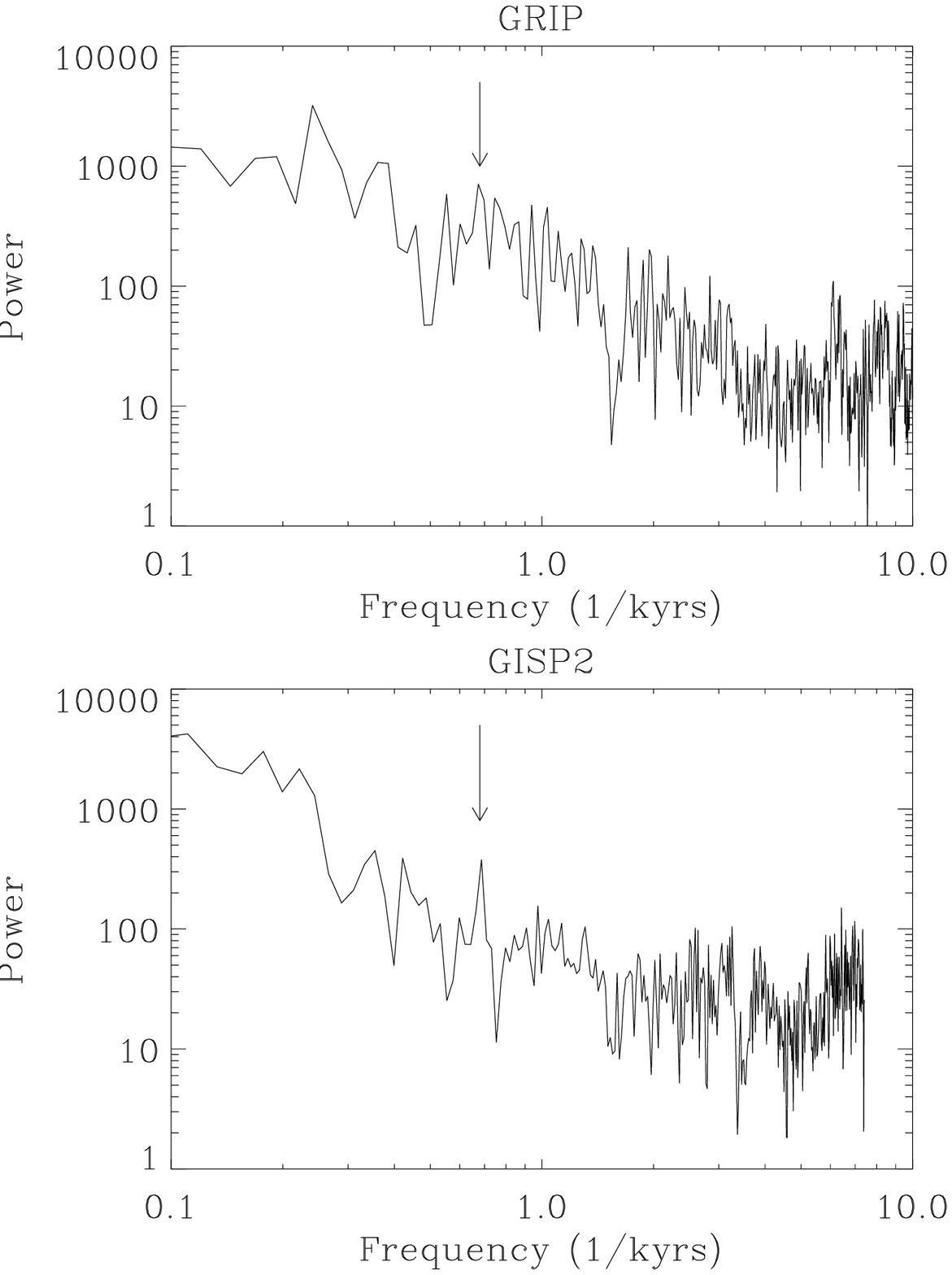}
\caption[]{}
\label{powerspectra}
\end{center}
\end{figure}

The power spectra of the GRIP and GISP2 records (figure 
\ref{powerspectra}) are estimated from the irregularly spaced data using the 'Redfit 3.5' 
routine \cite{schulz:2002a}.
The appearance of the 1470-year peak in the spectrum from the GISP2 data (figure \ref{powerspectra}) is 
caused by the 
regular spacing of three consecutive DO events (numbered 5,6,7) 
\cite{schulz:2002}. 
This regular spacing is not nearly as pronounced in the GRIP icecore using the 
ss09-sea timescale \cite{johnsen:1997a}, 
explaining why the spectral peak is not as 
significant in the GRIP spectrum. 
The discrepancy will hopefully be resolved in the near future where a more 
precise
dating of the NGRIP icecore based on annual layers counting far back in the 
glacial period
is expected.
Levels of significance are not included since these are strongly depending on the 
null-hypothesis noise spectrum
assumed. The regular Ornstein-Uhlenbeck process usually 
assumed \cite{schulz:2002a} is not relevant, since the isotope record is far from 
this process.
Here we shall not argue for either the GISP2 or GRIP dating, although we note that 
one cannot favor one in
comparison to the other solely based 
on the strength of the 1470-year spectral peak as was done by Rahmstorf 
\shortcite{rahmstorf:2003}. 

\begin{figure}[!H]
\begin{center}
\epsfxsize=\textwidth
\epsffile{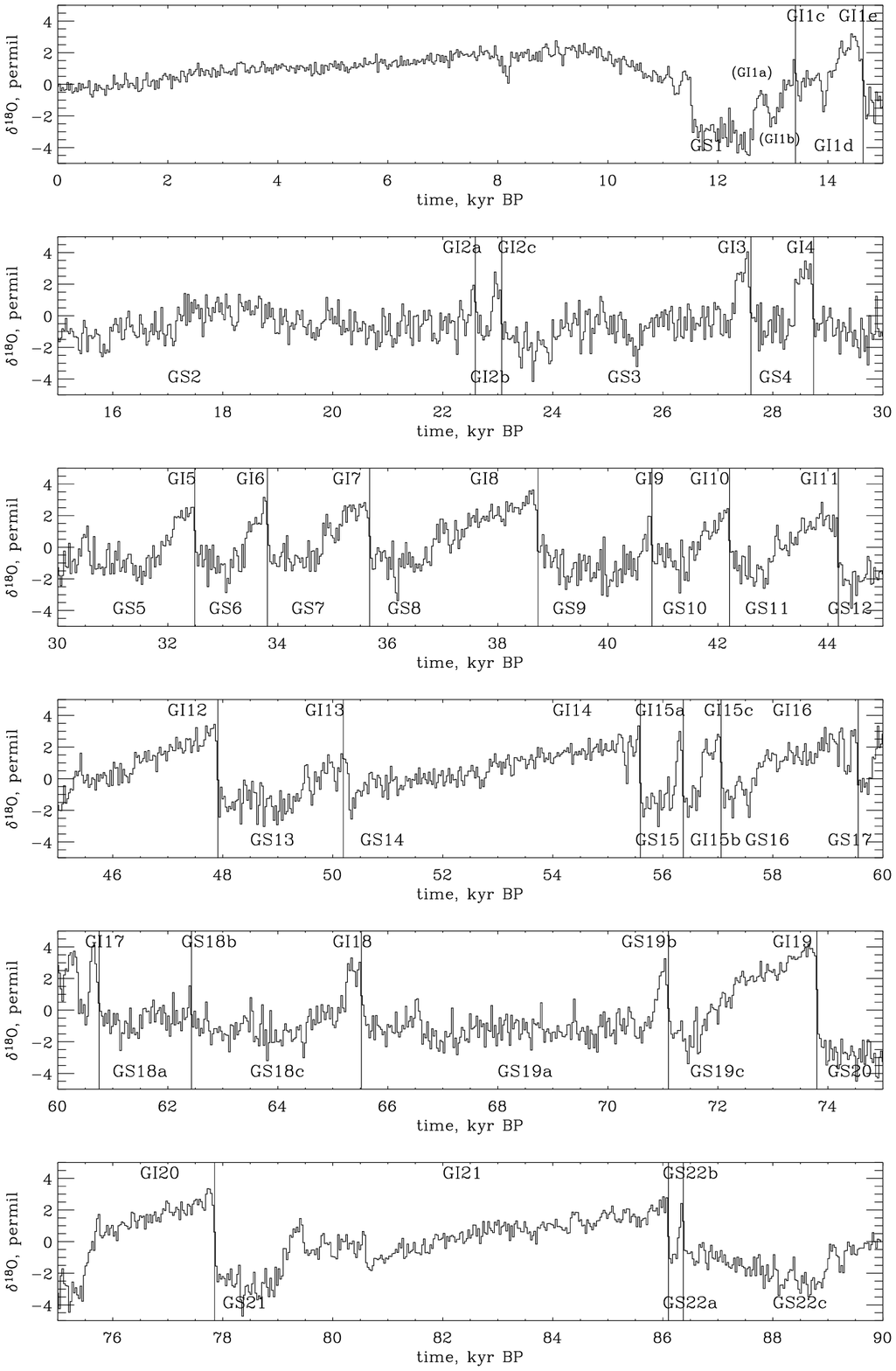}
\caption[]{}
\label{newDO}
\end{center}
\end{figure}

Besides the uncertainty in dating, some discrepancy exists in defining the 
warming events. 
The 'canonical' DO events \cite{dansgaard:1993} (figure 
\ref{newDO}) 
are based on visual inspection. Alley et al.  \shortcite {alley:2001} used 
a filtering procedure and threshold 
levels were adopted which produced 43 warming events. Schulz \shortcite{schulz:2002} found
that DO 
events 9, 15, 16 fell 
below the chosen threshold level, while the event around 65 kyr BP between 
events 18 and 19 was 
included. Rahmstorf \shortcite{rahmstorf:2003} only analyzed events after 50 kyr BP. 
Here event 9 was 
omitted while it was argued for an event ('A') prior to the Younger Dryas (YD)
and the termination of YD was included. We denote his event 'A' as 'GI1c' in accordance with
Walker et al. \shortcite{walker:1999}. 
In the following we will argue for a 
procedure of 
defining the warming events. The problem is two-fold: A spectral 
filtering must 
be decided and threshold values and a procedure for threshold crossings 
must be chosen. 

\subsection{The filtering}
There is a general consensus that a multi-millennial high-pass filter must be 
applied in order to eliminate the variations due to the orbital forcing. Whether 
this is a spectral filter or subtraction of a long-term running mean is not 
significant. It is with respect to this long term mean that the anomaly is defined.
The important problem is to decide for a low-pass in the other end 
of the spectrum. Most analyses have been performed on the approximately 200-year 
averaged isotope data available from the web archives. However, the spectral 
power in the timescale range shorter than 200 years is substantial. The effect 
of smoothing is that the positive extreme of a short timescale warming event 
will be reduced and possibly brought below a chosen threshold level. Thus analyzing a smooth 
signal introduces a bias towards omitting very short DO events. 

\begin{figure}[!H]
\begin{center}
\epsfxsize=85mm 
\epsffile{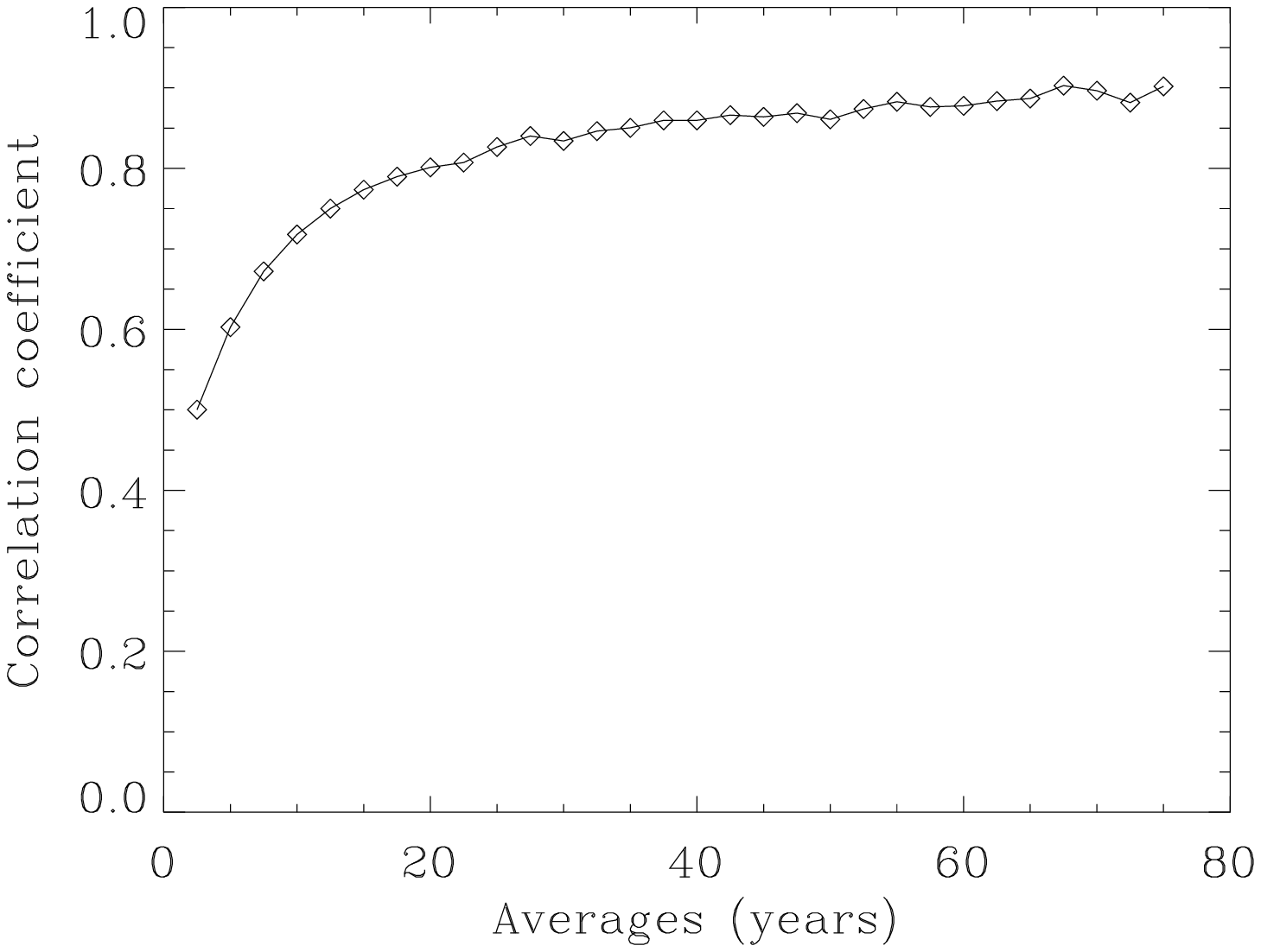}
\caption[]{}
\label{gripngrip}
\end{center}
\end{figure}

The shortest timescale 
fluctuations in the isotope records are dominated by glaciological noise \cite{ditlevsen:2003b} 
and local fluctuations. The low-pass should thus be determined such that the 
noise is reduced to an insignificant level in comparison to the true climatic
fluctuations. By comparing the high-resolution GRIP and NGRIP records we can 
estimate the noise level. Figure \ref{gripngrip} shows the correlation between 
the GRIP and the NGRIP records in the 27 - 37 kyr BP period, where the NGRIP 
signal has been dated by matching the DO events 3-7 
to the GRIP-ss09-sea timescale. The correlation 
coefficient is plotted as a function of the low-pass filter on both signals. The 
correlation between the raw high-resolution signals is approximately 0.5, while 
for the 30-year low-passes the correlation has increased to more than 0.8. The 
two core drilling sites are located more than 300 km apart so we conclude that the 
30-year low-pass signal is representative for climate fluctuations and
thus appropriate for the analysis.

\subsection{The threshold crossing}
\begin{figure}[!H]
\begin{center}
\epsfxsize=85mm 
\epsffile{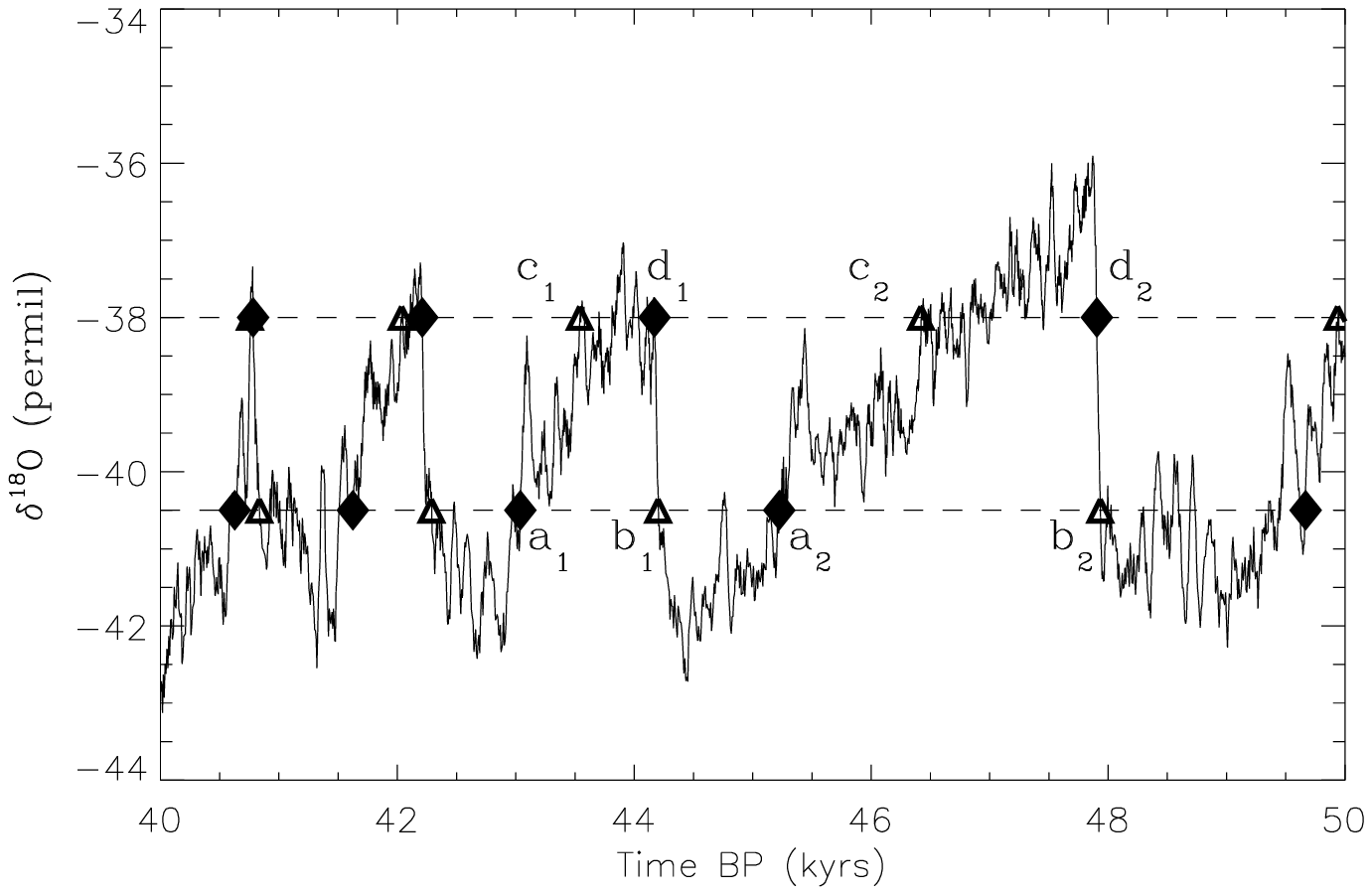}
\caption[]{}
\label{updown}
\end{center}
\end{figure}

The two distinct quasi-stable climate states are most clearly seen in the bimodal
distribution of the dust-calcium signal \cite{ditlevsen:1999a}, but is also
apparent in the $\delta^{18}O$ signal where the warm interstadial states are 'sawtooth shaped',
characterized by gradual decreases in the signal prior to jumping into the stadial state \cite{alley:1998,schulz:2002b}. Accordingly, the signal can be split into these two states and two 
transition states, cold to warm and warm to cold \cite{ditlevsen:2003b}. The initiation of the warm 
state is defined as the first up-crossing of the high threshold following an up-crossing 
of the low threshold. Similarly the initiation of the cold state is 
defined from the first down-crossing of the lower level following a down-crossing 
of the upper level (figure \ref{updown}). The terminations are defined as the last up (down) 
crossings of the lower (higher) level prior to an up (down) crossing of the 
higher (lower) level. The periods between terminations and initiations are the 
transition periods. 

The reason for this definition is that if the signal fluctuates around the higher
(lower) level in one of the climatic states, this does not lead to a jump unless
this follows a period where the signal was below (above) the lower (higher) level. 
This definition is advantageous in 
comparison to the definition used by Schulz \shortcite{schulz:2002}. Here the signal is 
defined to be within the warm state whenever the value of the anomaly signal is above 
the higher threshold. This has the consequence that a warm period can 
more easily be split into more periods as perhaps happens for DO event 17
\cite{schulz:2002}. The definitions applied here clearly identify whether DO 
event 17 should be regarded as one or two events. In our analysis it remains one DO event (GI17).

In the following the lower threshold is chosen as -1.0 permil anomaly and the 
higher threshold as +1.5 permil anomaly from the 10-kyr high-pass isotope 
signal. The asymmetry is because  
the climate system persists in the cold state longer than in the warm state.
The result is shown in figure \ref{newDO}. The vertical full lines indicate the
first up-crossings into the warm states. The result is quite robust in the
sense that the jumping times are rather insensitive to the threshold values chosen.

\section{The waiting time distribution}
\begin{figure}[!H]
\begin{center}
\epsfxsize=85mm 
\epsffile{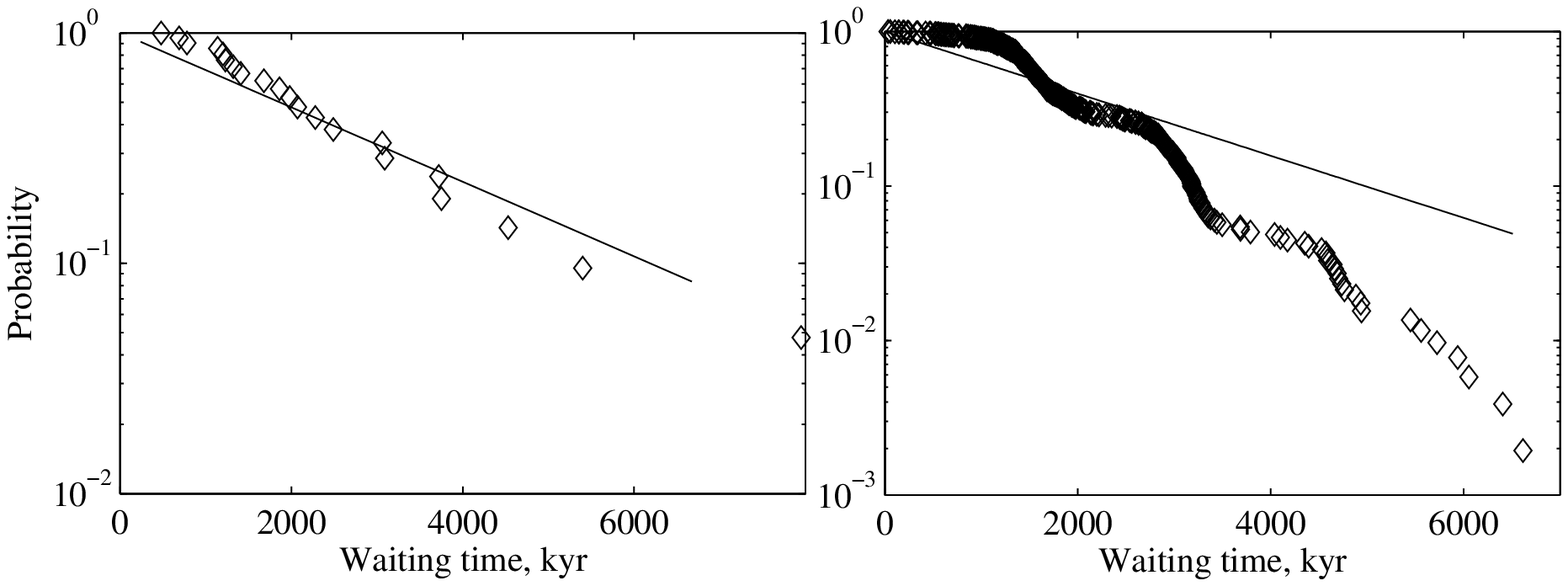}
\caption[]{}
\label{distribution}
\end{center}
\end{figure}

The transitions from the cold to the warm state are much more abrupt than the 
opposite transitions, so the waiting times between consecutive up-crossings are 
less sensitive to the choices of filter frequencies and thresholds and are thus 
used for the further analysis. The cumulated distribution of waiting times is 
plotted in figure \ref{distribution}, left panel. From a visual inspection the observed 
distribution seems to be sampled from a Poisson process, which has an 
exponential distribution. The mean waiting time $t_m$ is 2.8 kyr, which is defined by the
cumulated exponential distribution $P(t)=1-\exp(-t/t_m)$ shown as the straight line. 
The discrepancy with 
the earlier findings \cite{alley:2001,schulz:2002,rahmstorf:2003} lies mainly 
in the inclusion of the shortest warming events. Whether or not the doubtful event 'GS18b'
is included does not change the waiting time distribution significantly and does not
alter the following statistical analysis.    

\section{An upper bound on the periodic component}
Even though the distribution shown in figure \ref{distribution}, left panel, disfavors the 
previously claimed periodic component, the record is relatively short and, as 
Alley et al. \shortcite{alley:2001} noted, could be a realization of different possible 
processes.  
The waiting time distribution for a stochastic resonance process has a step-like 
structure with the first big step around the period and with exponentially 
smaller steps for multiples of the period (figure \ref{distribution}, right panel). A way of quantifying the degree of
periodicity is then to calculate the root mean square difference between the 
distribution and the best-fit exponential distribution. Assuming the 
existence of a periodic component with period of 1470 years, we can then estimate the 
strength of the periodic forcing in comparison to the barrier for purely
noise induced transition. Within 
the framework of the stochastic resonance model this amounts to investigating 
the strength of the resonance.

We have generated a series of data from a stochastic resonance model with the 
same mean waiting time as observed in the isotope signal but with different 
strength of the periodic component. For each realization of 
length similar to the length of the isotope signal we calculate the root mean 
square difference from the exponential distribution. The stochastic resonance 
model is described by the non-autonomous Langevin equation;
   
\begin{equation}
\frac{dT}{dt}=F(T)+A \cos(2\pi t/\tau)+\sigma \eta(t).
\label{model}
\end{equation}
The first term represents the drift with two stable states separated by a
potential barrier. The drift is derived from a potential $F=-dU/dT$, with
$U(T)=T^4-a_3T^3-a_2T^2+a_1T$, where $a_1,a_2,a_3$ are constants \cite{cessi:1994,ditlevsen:1999a}. 
The second term in (\ref{model}) is the periodic component with period $\tau$ 
and amplitude $A$. The third term is a white noise forcing with intensity $\sigma$.

For $|A|\ge A_c,~~A_c=-a_1-a_2a_3/2-a_3^3/8+8(a_2/6-a_3^2/16)^{3/2}$ the 
system will jump between the two stable states 
through a hysteresis loop periodically, while for $|A|< A_c$ the 
bifurcation points are not reached and the jumping must be noise induced.    

\begin{figure}[!H]
\begin{center}
\epsfxsize=85mm 
\epsffile{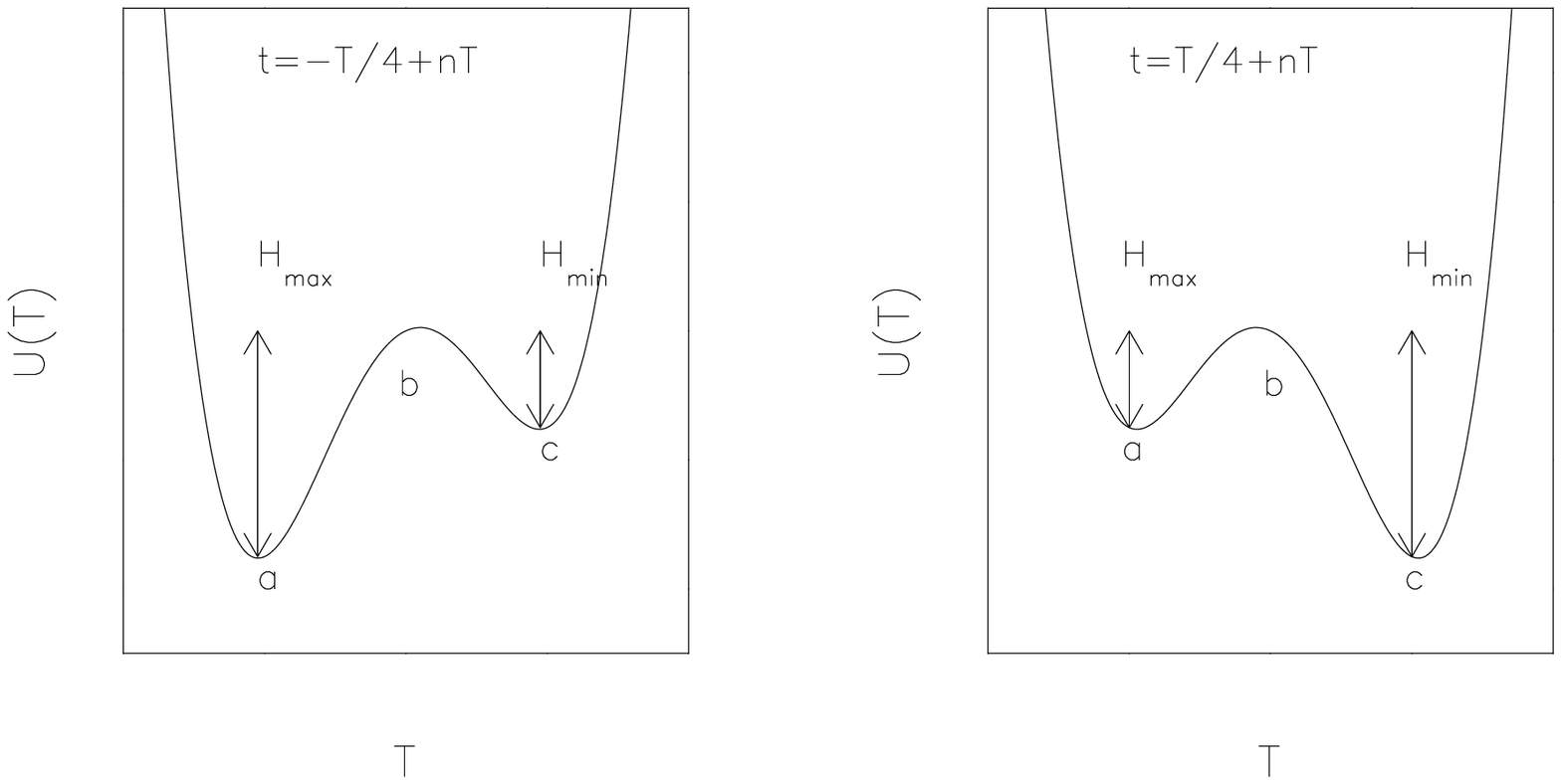}
\caption[]{}
\label{potential}
\end{center}
\end{figure}

A schematic of the potential and the periodic forcing is shown in 
figure \ref{potential}.
Disregarding non-exponential prefactors, the Kramer waiting time for pene\-trating 
the barrier going from the well 
containing the stable state $a$ ($c$) to the well containing the stable state $c$ 
($a$) is estimated $T_{ac}\sim exp(H_{min}/\sigma^2)$
($T_{ca}\sim \exp(H_{max}/\sigma^2)$) where $H_{min}$ ($H_{max}$) is the height 
of the barrier \cite{gardiner:1985}. The criterion for the noise intensity to 
obtain stochastic resonance is 
\begin{equation}T_{ac}<<\tau<<T_{ca}.\label{tscale}\end{equation} 
The climate state will 
then with high probability jump
from state $a$ to state $c$; within time $\tau/2$ the potential has changed
due to the periodic component and the state will with high probability jump
from $c$ to $a$, and so on. 
 
\begin{figure}[!H]
\begin{center}
\epsfxsize=\textwidth
\epsffile{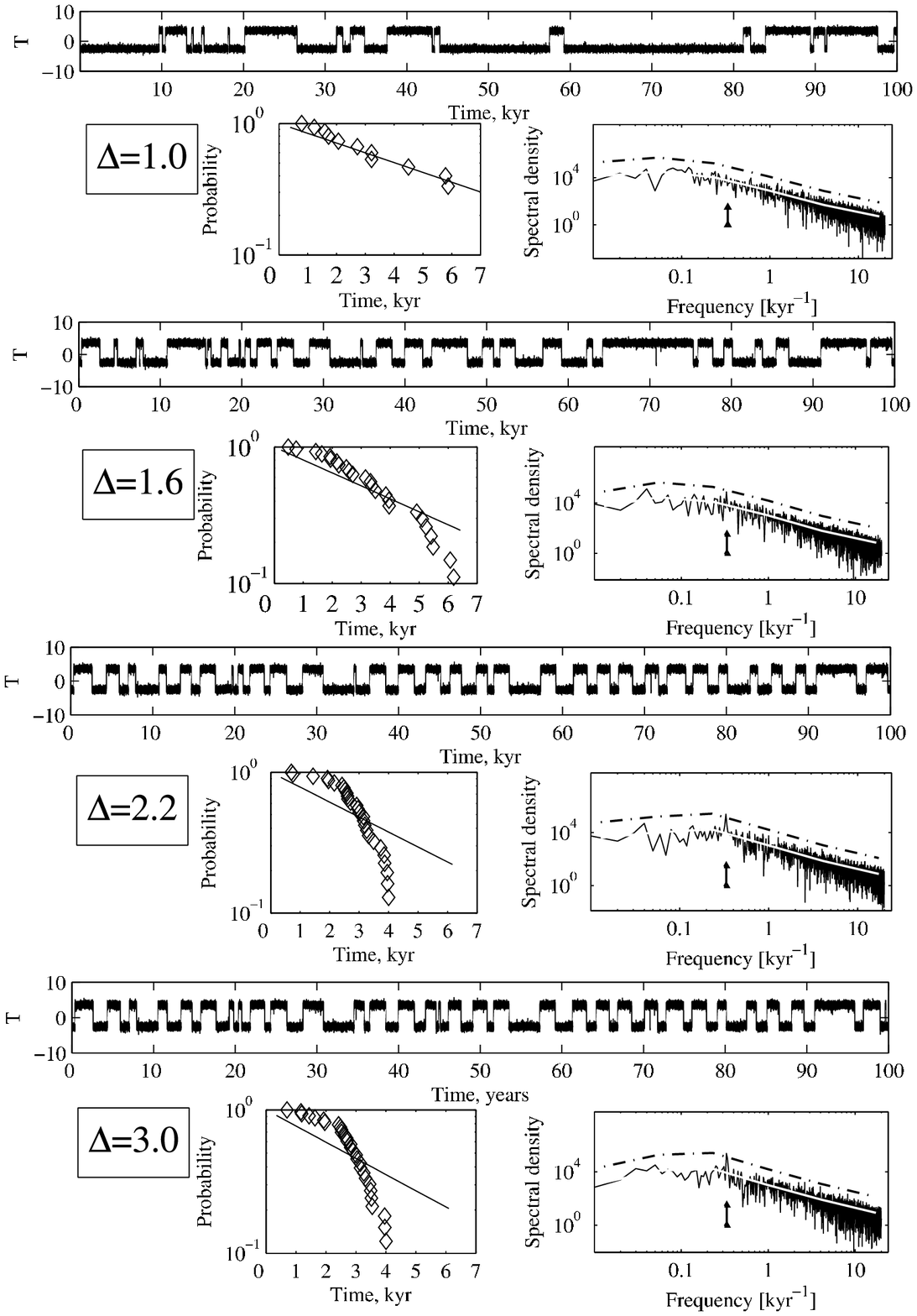}
\end{center}

\caption[]{}
\label{simulations}
\end{figure}

By varying the strength of the periodic component, still tuning the noise intensity to the 
stochastic resonance, we can obtain an estimate of the
amplitude of the periodic forcing component in comparison to the barrier in the system.
Figure \ref{simulations} shows a set of realizations of equation (\ref{model}) for 
different strengths of the periodic forcing. Each realization is represented in a
set of three figures; long-top -, left - and right panels.
The strength of the stochastic resonance in the system is determined by the 
separation in timescales expressed in equation (\ref{tscale}). When the  
noise is tuned to the resonance the criterion is $\Delta\equiv T_{ca}/T_{ac}\sim 
\exp[(H_{max}-H_{min})/\sigma^2]>>1.$ 
The four panels correspond to $\Delta = 1.0, 1.8, 2.5, 3.4$ respectively. 

\begin{figure}[!H]
\begin{center}
\epsfxsize=85mm 
\epsffile{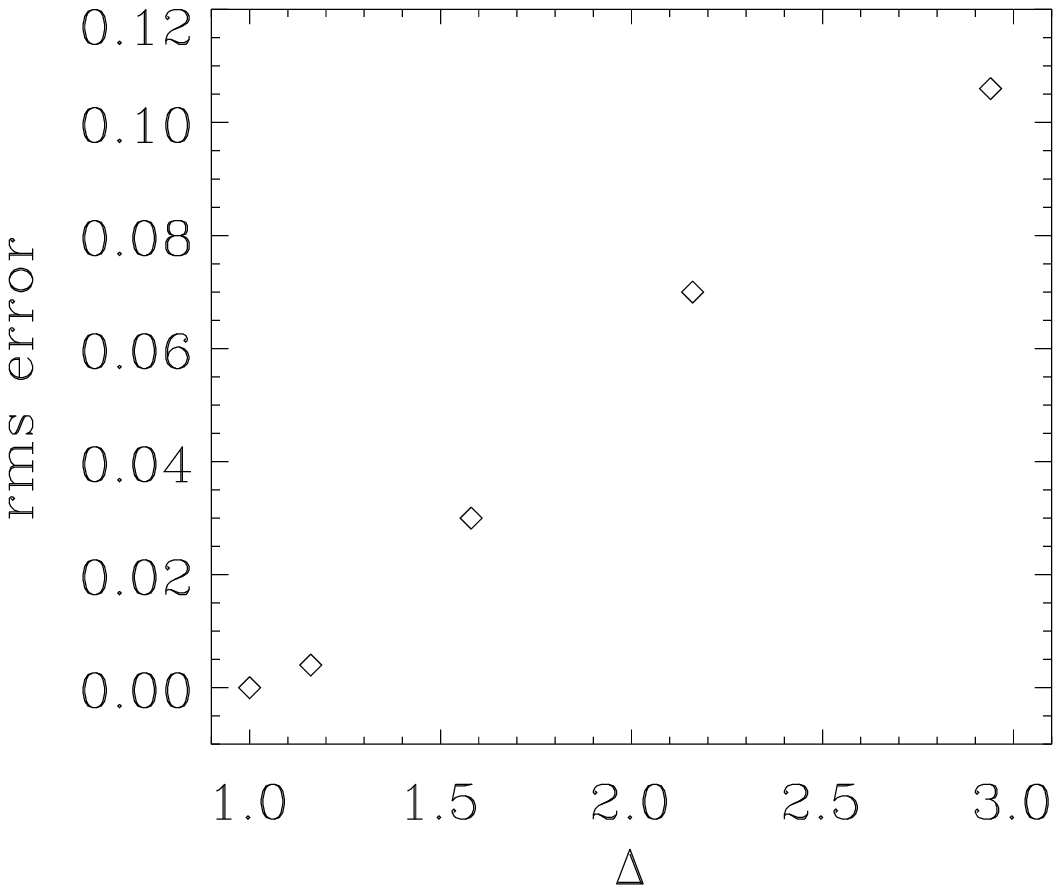}
\caption[]{}
\label{rms}
\end{center}
\end{figure}

Going from 
top to bottom we see that the periodic component emerges. In the right panels the
spectral peak at $f=2\pi /\tau$ emerges and exceeds the 99\% confidence level \cite{crowley:1986} for $\Delta > 2.5$. 
The cumulated waitingtime distributions are shown in the left panels. The step-like structure from the
periodicity emerges as $\Delta$ increases.
The top set of panels ($\Delta =1.0$) shows the pure Poisson process. 
A quantitative measure of 
the deviation from the Poisson process is the root mean square (rms) distance of the 
waiting time distribution from an exponential. This is similar to the measure used for the
Kolmogorov-Smirnov test which
can not be applied when we obtain the best fit distribution from the data. 
In the following this rms is denoted the 
'error'. For each of the models 
shown in figure \ref{simulations}  the 'error' is calculated from a sufficiently long
simulation (figure \ref{rms}). The accuracy is within the size of the plotting symbols. 

\begin{figure}[!H]
\begin{center}
\epsfxsize=120mm 
\epsffile{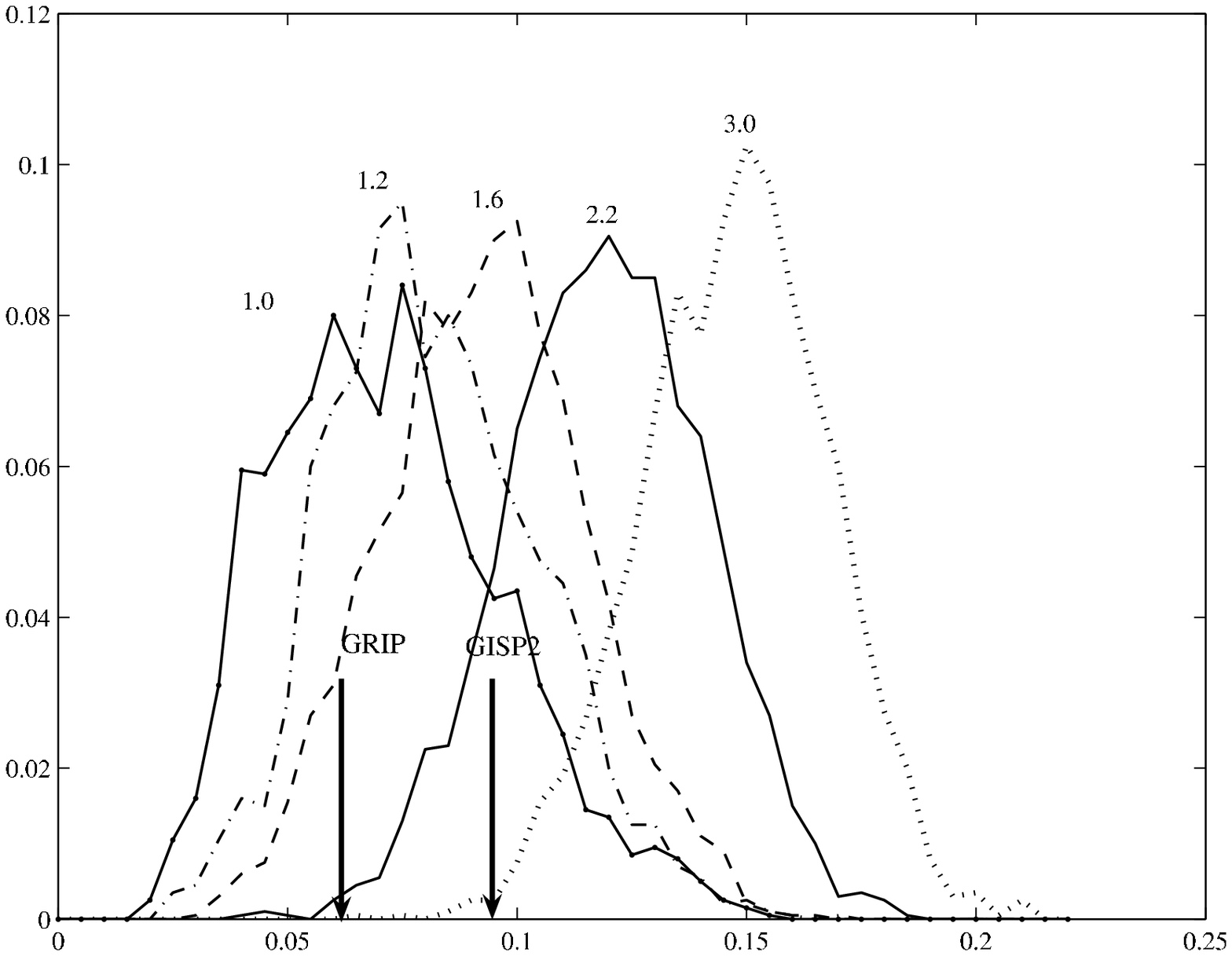}
\caption[]{}
\label{errordistributions}
\end{center}
\end{figure}

In order to evaluate the data series,
we have simulated a large ensemble of time series generated by eq. (\ref{model}) of 
same length as the isotope record for a given noise intensity. For each 
realization we calculate the root mean square distance to the exponential 
distribution. By this procedure the distribution of 'errors' is calculated.
In this way the maximum likelihood model for generating the observed record is 
obtained. Figure \ref{errordistributions} shows the error distributions for the 
models generating the series shown in figure \ref{simulations}. The arrows 
represent the error obtained from the GRIP and GISP2 isotope records. The maximum likelihood 
model has $\Delta = 1$ for the GRIP dating, while $\Delta=1.6$ for the GISP2 dating. For the  
model with $\Delta=2.2$, 2\% of the realizations have an error less than the one
measured for the isotope signal. We thus reject the hypothesis that the data are generated by a process
with $\Delta \ge 2.2$ with 98\% confidence for the GRIP dating. Correspondingly, we
reject a process with $\Delta \ge 3.0$ with 98\% confidence for the GISP2 dating.
The NGRIP record is presently dated by fitting to the GRIP timescale and is thus not independent.  

\section{Summary}
The isotope record shows the jumping between the warm DO 
interstadial state and the cold stadial state with waiting times in the 
millennial timescale range. We have used an objective procedure based on 
correlating the GRIP and NGRIP records to decide the resolution of the climate 
record. A high and a low threshold for the anomalies are applied. The 
transitions are defined from consecutive level crossings. By this procedure we 
obtain the 'canonical' DO events and an additional set of short events. This
happens either by
splitting some of the 'canonical' events (1, 2, 15) or by defining 
events previously in the cold periods (18, 22). Assuming the record to be 
generated by a dynamics described by equation (\ref{model}) the strength of a periodic 
component in the forcing in comparison to the strength of the barrier is expressed
through the non-dimensional parameter $\Delta$.
It will be relevant to 
extend the present analysis to include the oldest 
DO events from event 22 to the newly discovered event 25 \cite{ngrip:2004} when a reliable 
dating is obtained for the NGRIP record.  
The observed record is highly probable as 
a realization of a purely noise driven process without a periodic component.
This finding suggests that a mechanism for the sudden shifts between the
two distinct climate states involving changes in the North Atlantic deep water formation.
The shifts could be results of the erratic fluctuations in fresh water formation and
heat trasfer to the ocean surface. These fluctuations were stongest in the 
last glacial maximum where there is a tendency for more frequent shifts \cite{rahmstorf:2003,ditlevsen:1996b}. The
occurence of event 25 shortly after the termination of the Eem period, before
substantial icevolumes have build up, suggests that the same climate modes 
exist in the interglacial periods where the 'warm' mode persists \cite{ganopolski:2001}. 
The reason why we do not experience DO events (except perhaps the 8.2 kyr event) 
in the Holocene climate could be
the low intensity of fluctuations rather than the stability of the 'warm' mode versus
the 'cold' mode as suggested by Ganopolski and Rahmstorf \shortcite{ganopolski:2001}.

{\bf Acknowledgement} Discussions with S. J. Johnsen is greatly appreciated. KKA thanks Carlsberg for funding.


\newpage
\bibliographystyle{authordate1}
\bibliography{/disk3/pditlev/documents/shell_model/full/climate}


\newpage
\begin{center}
FIGURE CAPTIONS
\end{center}
\newcounter{fig}
\begin{list}{Fig. \arabic{fig}}
{\usecounter{fig}\setlength{\labelwidth}{2cm}\setlength{\labelsep}{3mm}}

\item
The power spectra of the GRIP and GISP2 icecores. The difference in 
the spectra is
due to the discrepancy in dating. The 1470 years peak indicated by the arrows is 
only pronounced 
in the GISP2 record.

\item
The 10 kyr high-pass of the 30 years averaged GRIP $\delta^{18}O$ 
record. The numbering 
is the 'canonical' dating from Dansgaard et al. (1993). The notation Greenland
Interstadial (GI) and Greenland Stadial (GS) is adapted from Walker et al. (1999).
The numbering followed by a letter
is the DO events obtained using the procedure described in the text. Again following
Walker et al. (1999), when a
previously unsplit interstadial is split into a sequence of shorter interstadial/stadial periods
it keeps its assigned 'GI' preamble. The periods are then assigned letters 'a,b,c,...' after
their numbering, where 'a' is the youngest. This means that fx. 'GI15b' is a stadial state
despite the 'GI' preamble. With similar convention the stadial states (GS18, GS22) are split. 
One exception from our convention is event GI1, which is split according to Bj{\"o}rk et al. (1998).
Only events GI1c and GI1e qualify as DO events.
The anomaly
levels used for the up- and down-crossings are +1.5 permil and -1.0 permil respectively.
The short event 'GS18b' at 62.4 kyr BP appears with this choice of anomaly levels even
though it is doubtful whether it qualifies as a DO 
event. This cannot be determined solely based on the icecore records. Ideally it
should be found in deep sea cores, or other records indicating if it can be related
to a change in the THC.  

\item
The correlation between the GRIP and the NGRIP records as a function 
of the low-pass. The
correlation falls off for fluctuations on timescales shorter than about 
30 years. For the 
short timescales the glaciological noise and local fluctuations dominate, while 
the 30 years low-pass
is taken to define a robust climate signal.

\item
The separation between stadial -- and interstadial states is done on
the 30 years low-pass of the isotope signal.
The periods are defined by the separation points being the first down-crossing 
time
of a lower level (lower dashed line) after the signal has crossed up through an
upper level (upper dashed line). The upper level is the 1.5 permil anomaly while the 
lower level is 1.0 permil negative anomaly. 
This defines the points marked with $a$'s and $d$'s (diamonds)
The $b$ and $c$ points (triangles) are obtained in the same way by
moving backward in time (from left to right in the plot) \cite{ditlevsen:2003b}.

\item
Left panel: The cumulated waiting time distribution between consecutive up-crossings into 
warm events. The probability
for waiting more than $\tau$ is plotted as a function of $\tau$. The straight 
line is an exponential distribution
with mean waiting time of 2.8 kyr. Right panel: The cumulated waiting time distribution for a
long realization of a stochastic resonance. The step-like structure deviation from a
straight line exponential distribution reflects the periodicity in the signal.

\item
The potential in the two extremal positions for $t=\pm\tau/4$ 

\item
The long panels show realizations of equation (1) with $\Delta = 1.0, 1.6, 2.5, 3.4$. $\Delta$ is the dimensionless
ratio of expected waiting times for jumping from the shallow -- to the deep -- and jumping from the deep -- to the 
shallow well, see text for explanation. The corresponding 
power spectra are shown in the following right panels. The 99\% significance levels is defined by the a posteriory 
criterion ({\it Crowley et al.,1986}). Using this criterion, the spectral peak at $\tau^{-1}$ is significant 
for $\Delta\ge 2.5$. The cumulated waiting time distributions are shown in the left panels, going from the purely 
noise driven Poisson process towards an almost periodicly shifting
signal.

\item
The root-mean-square (rms) errors expressing the deviation from an exponential
distribution for the simulations shown in figure \ref{simulations}. The rms errors are
calculated from simulations much longer than the ones shown in figure \ref{simulations}

\item
The distributions of calculated rms errors from realizations of length
corresponding to the length of the observed record. The arrow shows the value obtained
for the isotope signal. This value is most probably generated by a process with $\Delta=1.0$ 
which is the purely noise driven process.

\end{list}

\end{document}